\documentclass[conference]{IEEEtran}

\usepackage{mathptmx}       % selects Times Roman as basic font
\usepackage{helvet}         % selects Helvetica as sans-serif font
\usepackage{courier}        % selects Courier as typewriter font
\usepackage{type1cm}        % activate if the above 3 fonts are
\usepackage{makeidx}         % allows index generation
\usepackage{graphicx}        % standard LaTeX graphics tool
\usepackage{multicol}        % used for the two-column index
\usepackage[bottom]{footmisc}% places footnotes at page bottom

\usepackage{cite}
\usepackage{amsmath,amssymb,amsfonts}
\usepackage{algorithmic}
\usepackage{graphicx}
\usepackage{textcomp}

\begin{document}

%\runningheads{C.~C.~M}{Self-organizing Life Cycle Management of Mobile Ad hoc Networks}

%\articletype{RESEARCH ARTICLE}

\title{Self-organizing Life Cycle Management of Mobile Ad hoc Networks}

\author{\IEEEauthorblockN{C. Caballero-Gil, P. Caballero-Gil, J. Molina-Gil}
\IEEEauthorblockA{Department of Statistics, Operations Research and Computing, \\
University of La Laguna,
Spain\\
Email: \{ccabgil, pcaballe, jmmolina\}@ull.es }}
\maketitle
\begin{abstract}
A Mobile Ad hoc NETwork (MANET) is a type of wireless network without any infrastructure, where nodes must adapt to the changing dynamic situations that result from their mobility. Because of the decentralization of nodes and the security needs of communications, management of MANETs must be self-organized, which is a major research challenge. In order to cope with the intrinsic properties of MANETs, a new decentralized management system for MANETs called Self-organizing Life Cycle Management (SLCM), is here fully described and evaluated. Regarding security, node authentication is the most critical component of access control in any network, and in particular in MANETs. Broadcasting is also a fundamental data dissemination mechanism in these networks. Both aspects have received special attention when defining the proposed SLCM system. In particular, both a strong access control algorithm based on the cryptographic paradigm of zero-knowledge proofs, and a three-step broadcast protocol, are here defined. This work includes the performance evaluation of the scheme, and the obtained experimental results show that SLCM significantly improves both the quality and the security of life cycle management of self-organized MANETs.
\end{abstract}

%\keywords{MANET; Broadcast; Strong authentication; Access control; ZKP}

\section{Introduction}
The main aim of the next generation networks \cite{Prahmkaew, Xie} is the offer of ubiquitous connectivity services to mobile users through heterogeneous wireless networks such as MANETs, which are nowadays being used in many application scenarios, including military, emergency, health, environmental, surveillance, education, business, commercial, etc. The wide spread of these networks is mainly due to the fact that they can be built quickly because they do not need any fixed or centralized infrastructure such as central servers, base stations or access points. 

Unlike the architecture of other networks, in a MANET every node may work as a host and router at the same time, and each movement of a mobile node affects the topology of the network. Thus, routing in ad hoc networks is one of the most extensively studied problems \cite{Chang, Imani} In addition, the lack of central infrastructure also makes it difficult, if not impossible, the existence of an authority that manages the operation and security of the network. This paper has been prepared with the aim of solving these problems.

Resource constraints and, in particular, limitations in communication and computation, gradual deployment and need of scalability, lack of central or fixed infrastructure, and unreliability of the radio media are some of the main challenges that must be taken into account when designing any protocol for MANETs. \cite{Bonnefoi} gives a useful state of the art on practical and global solutions for MANET deployment. Regarding security, four  important aspects are authenticity, confidentiality, integrity and privacy. Among them, authentication can be considered the most critical one because it enables the proper identification of legitimate nodes, allowing the fulfillment of any other security service.

The most widespread approach to authentication in networks is based on weak schemes formed by maximum-disclosure proofs with secret time-invariant passwords. Their major security concern is possible eavesdropping and subsequent replay of secret passwords. Two well-known solutions to this security problem are variable passwords and minimum-disclosure proofs. The authentication protocol here proposed combines both concepts in order to define a strong authentication scheme specifically thought for MANETs.

In this paper, we focus on monitoring and configuration management in MANETs, that is to say, on the processes to control nodes and data in these networks in order to maximize their security. In particular here we describe every phase of the proposed life cycle management scheme \cite{Badonnel}. A basic requirement when configuring MANETs lies on the self-organizing ability of network nodes. Recently, several self-management mechanisms for MANETs have been proposed in the bibliography for different actions such as path discovery \cite{Kambourakis} or clustering \cite{Yang}. 

Regarding authentication, numerous protocols have been specifically developed for MANETs. \cite{Aboudagga} motivates the need for specific authentication management architectures for MANETs, which is one of the main issues of the present paper. The simplest solution for MANETs is location-limited authentication, which is based on that most ad-hoc networks exist in small areas and physical authentication may be carried out between nodes that are close to each other \cite{Stajano}. However, such solutions are not useful for the general case. Short-lived MANETs are specifically analyzed in \cite{Saxena}, where a secure, efficient and non-interactive access control protocol was proposed.

Other authentication proposals \cite{Hahm} are based on public key cryptography, what leads to the problem of public key certification. In general, the typical approach to this issue is through the existence of a Certification Authority (CA) that guarantees the validity of all node identities. In the case of MANETs, such a role can be played by a distributed group of nodes \cite{Luo, Zhou}. This approach can degrade the CA availability. Another solution to the certification problem in MANETs is based on the chain of trust paradigm \cite{Hubaux}. Its main problem is the danger that an attacker can control the signing process by compromising only a small number of nodes.

In \cite{Tseng} a group key agreement protocol for mobile nodes is proposed, where all participants cooperatively establish a group key to encrypt messages. Such a work provides simulation data, like \cite{CCMQ}, but the shown results are only described in a superficial way, with no images. Other schemes based on threshold cryptography and ring signatures are surveyed in \cite{Azer, Freudiger} but the same situation is repeated. 

In the present work, a legitimate node presents its credentials to another legitimate node in an attempt to access the network according to an authentication process based on the established cryptographic paradigm of Zero-Knowledge Proofs (ZKPs), which were introduced in \cite{Goldreich}. Until now a few publications have mentioned the proposal of authentication systems for MANETs using ZKPs \cite{Caballero}, and none of them includes the authentication proposal in the context of a whole life cycle management dealing with the related problem of topology changes due to mobility, which is exactly the main objective of  this work. 

The rest of this paper is organized as follows. Section \ref{goals} discusses the security motivation of this research. Section \ref{preliminaries} presents preliminaries including general aspects and notation, and a new optimized way to perform broadcast in MANETs. Section \ref{scheme} shows a complete description of the operation of the proposed scheme called SLCM, providing specific details of every phase: network initialization, node insertion, access control, proofs of life and node deletion. The security of the proposed scheme is discussed in Section \ref{analysis} while Section \ref{evaluation} illustrates its performance analysis. Finally, some conclusions and open questions complete the paper.

\section{Security Goals}
\label{goals}

Efficiency, reliability and security are our main design goals for the self-organizing life cycle management that we propose in this work for MANETs. In order to describe the security objectives, we distinguish between outsider and insider nodes. 

An outsider node is a node that is not an authorized member of the MANET whereas an insider node is an authorized legitimate member of the network. The security goal of this research is to develop mechanisms that protect a self-organized MANET without any central authority against malicious behavior from outsider nodes as well as from insider nodes. 

Detecting attacks from insiders is one of the tasks of Intrusion Detection Systems (IDSs).  Since insiders have access to the MANET, it is easy for them to launch sophisticated attacks. In this paper we propose a response system providing the capability to effectively cut off compromised insiders from the MANET. In addition, the scheme offers some level of protection against insiders who try to forge packets and impersonate other insiders.

We now describe our main security goals in defending the underlying network against outsider nodes. 

Any packet transmitted by an outsider node should be immediately dropped by the receiving insider node at the first hop with a very high probability. In other words, packets sent by outsiders should not be allowed to be propagated through the MANET. By fulfilling this requirement, we can successfully guard against a myriad of attacks launched by outsiders, such as DoS (Denial of Service), wormhole attacks, man-in-the-middle, SYN flooding, etc. This is because in this way we are effectively disabling the outsider's ability to route any packet to any node that is not its neighbor. However, the aforementionated requirement dictates that every packet has to be authenticated at every hop, which in turn means that the authentication mechanism should be extremely efficient.

On the other hand, the outsider node is assumed to have the capability to spoof its identity, data such as its IP and MAC addresses, so these are not considered reliable in the schema.

The outsider is also assumed to have access to the wireless channel so it can eavesdrop on legitimate traffic. Thus, if the traffic is supposed to remain confidential, end-to-end encryption should be used to protect it, and, in any case, legitimate traffic should not be useful to launch attacks.

If an IDS is used to discover a compromised insider, the system must be able to exclude it from propagating any packet within the network. Regarding this issue, Certificate Revocation Lists (CRLs) might be used to revoke the certificates of compromised insiders. These CRLs have to be sent to the whole network when a group of legitimate nodes detect a malicious node. Then, its CRL receiving every node update. If some node does not update its CRL because it was off-line, it can check it against the version of the CRL that is sent during the life cycle of the network. In order to check whether a node is compromised or not, nodes must verify the information sent by its neighbors. If multiple nodes receive incorrect or inconsistent information repeatedly from the same source, this group of nodes introduce the information of the suspicious node in the CRL. In this case there must be a minimum number of nodes that agree to sign the revoked node, which is a threshold that depends on the size of the MANET.

\section{Preliminaries} 
\label{preliminaries}

The proposed scheme is thought for small and medium-sized MANETs where security for communications is required at the expense of sending control packets. Since MANETs do not have any fixed infrastructure, their capacity to support network routing is limited so the schema is not appropriate for large networks as it would increase the complexity and the number of control packets too much.

Due to the absence of fixed infrastructure, routing is a hard problem in MANETs. The proposed scheme allows to know which nodes are authenticated and on-line, without any fixed infrastructure. In this paper we do not propose new routing schemes because the simulations of our proposal show that existing schemes such as DSR or OLSR give good results without saturating communications.

The following sections describe, respectively, a new optimized protocol for broadcast in MANETs that is used in the SLCM scheme, an overview of the SLCM proposal and a description of the used notation.

\subsection{Optimized Broadcast}

The broadcast protocol called \textsl{GRI} (Go-Return-Information) is a new optimized broadcast scheme designed to solve some problems in wireless communications without centralized authority. The protocol consists of three simple phases called go, return and information. 

In the first stage the node that initiates the GRI broadcast sends a signal through broadcast with a request-response to all nodes that are within the transmission range of the network and each node receiving this message, forwards it to its neighbors. In the return stage, the nodes that are farther from the node that initiated the GRI broadcast, that is to say, the nodes that do not have anyone else to send the message, start the return phase of the GRI broadcast. In this stage, nodes send their identifiers to the node that initiated the GRI broadcast. Nodes have a timer to wait for responses. So every node has to respond during that time to the emitter nodes. When the response goes through intermediate nodes, they add their identification to the response packet and forward it to the source node.

In the information phase the node that initiated the GRI broadcast gets all the information of the network and sends a broadcast again with all the information of the network to all nodes. With this simple protocol it is possible to control the entire network and to send relevant information to all nodes in the network by generating the least number of control packets in the network. 

\begin{figure}[htb]
  \centering
     \includegraphics[scale=0.32]{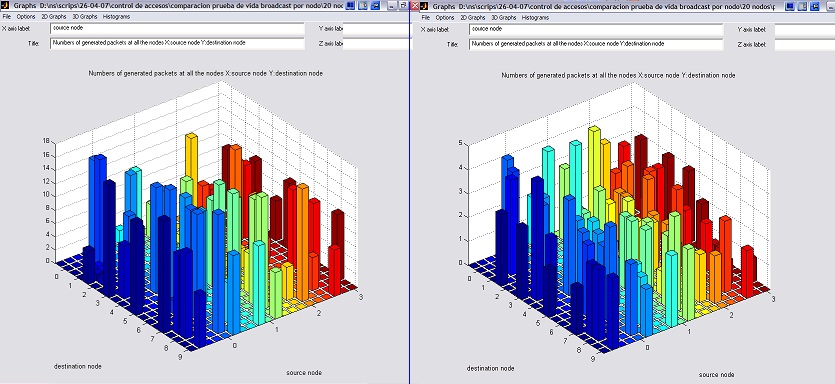}
  \caption{General vs. Optimised Broadcast}
  \label{fig:gri}
\end{figure}

Notice that in special situations such as for example if nodes are placed in a line, the packet size might grow quite large because if the initiator is the node on one end-point of the line, the return packet will contain the information gathered from all the nodes in the network.
As a visual example of the usual situation, Figure \ref{fig:gri} shows a comparison between packets generated by the tool trace. For 20 nodes the number of generated packets from each node by a normal broadcast and by the GRI broadcast are compared. Note that the number of broadcast packets generated by the GRI protocol is between 40\% and 60\% of the number of packets generated by the normal broadcast. Thus, the graph shows clearly that the results are better with the GRI broadcast than with the normal broadcast. 

\subsection{Outline of the Proposal}

The SLCM scheme presented here has been designed as an authentication scheme for membership in a group because when a node wants to become part of the network, it must be accepted by some legitimate nodes. The number of legitimate nodes required for the insertion of any node must be large enough to ensure that potential attackers captured by multiple nodes can not introduce new malicious nodes. This number depends on the size of the network.
 
According to the authors of \cite{Maki}, in any group management protocol it is necessary to establish robust methods to insert and delete nodes, and to allow access only to legitimate members of the group. For that reason, this work not only describes the procedure for controlling the access of legitimate nodes, but also the procedures for updating the network associated with insertions and deletions of nodes. In particular, in this paper the procedure for deciding which nodes are removed from the network is based on the time the node has been offline, so if a node has been offline for a long time (compared with a pre-arranged threshold parameter), it is removed from the network.

The cryptographic paradigm of Zero-Knowledge Proof forms the theoretical basis of the access control procedure described below. In particular, the protocol is applied for the particular case of the Hamiltonian Cycle Problem (HCP). A Hamiltonian cycle of a graph is a cycle that visits each vertex exactly once and returns to the starting vertex. The determination of whether there are cycles in a graph Hamiltonian is called the Hamiltonian Cycle Problem, which is an NP-complete problem. This problem was chosen for our design mainly because the upgrade of a solution due to an insertion or deletion of a vertex in the graph does not require a large computational effort. These operations are common in our application due to the high dynamism of the analyzed networks. However, similar schemes could be described on the basis of other NP-complete problems on graphs where the updating of a solution after the individual changes in the graph is also easy. This is the case of problems such as of Vertex Cover, Independent Set or Clique Problems, for example.

The proper performance of the proposed system is only possible thanks to the use of a chat application via the GRI broadcast scheme proposed above, since it makes possible for some legitimate online nodes can send a message to all online network nodes. The application allows publishing all information associated with the network upgrade. Although it is not necessary that the chat messages are transmitted secretly because they are useless for illegitimate nodes, since that information is necessary to update the authentication information, it is required that only online legitimate nodes can launch the GRI broadcast of the chat application.

All the data received through GRI broadcast of the chat application for an interval of time must be stored by each online node in a FIFO queue. These data allows the update of authentication information for all legitimate offline nodes whose access is authorized by the online nodes. The duration of this period, which will be denoted $T$, is an essential parameter because it indicates the maximum time allowed to be out of line for any legitimate node, and also the frequency of the proofs of life that will be described below. Consequently, this parameter must be agreed by all legitimate nodes.

\begin{figure}[htb]
  \centering
     \includegraphics[scale=0.47]{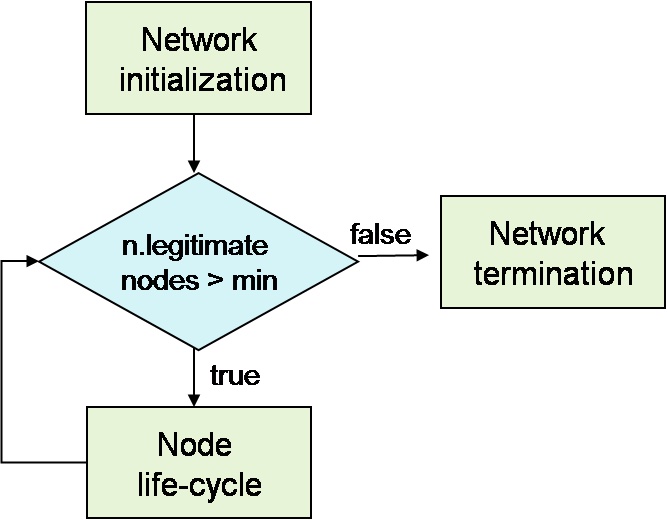}
  \caption{Network Life-Cycle}
  \label{fig:lifeCycle}
\end{figure}

The network life cycle has three main phases, as shown in Figure \ref{fig:lifeCycle}. Initialization is the first phase, where each member of the original network receives, either off-line or on-line, a piece of secret information playing the role of secret network key. The knowledge of such a secret network key can be used for access control to demonstrate eligibility of nodes in order to access protected resources or to offer some service to the network.

After the initialization phase, the legitimate nodes can participate in the network, so the node life cycle begins. 

Through the access control a legitimate node that has been offline proves its membership to a legitimate online node. In order to do it, the prover node must demonstrate its knowledge of the secret network key with a challenge-response scheme.

When a legitimate node is given permission to access the online status of the network, it has full access both to protected resources such as the GRI broadcast chat application, and to provide network services such as insertion of new nodes. 

The secret network key is continuously updated according to the changes in the network topology, so the secret key of a legitimate node expires if it is offline for too long. In that case, the node would have to be re-inserted in the network by a legitimate online node if it wants to join the network again. 

In our proposal, the secret network key is based on the difficulty of the HCP, so the number of legitimate nodes is a very influential parameter in such a difficulty. Thus, if the number of legitimate nodes decreases and becomes too small, the network termination is automatically performed, the life cycle of the network ends. 

Note that no adversary can steal any significant information, even though it accesses to all information sent through the GRI broadcast, or if it sniffs the data exchanged between a prover node and a verifier node through an access control procedure.

\subsection{Notation}

Notations used in the proposal are given below:

\begin{itemize}
\item $G_t=(V_t,E_t)$ denotes the undirected graph used at stage $t$ of the network life-cycle.

\item  $v_i \in V_t$ represents both a vertex of the graph and a legitimate node of the network.

\item $n=\left|V_{t}\right|$ is the order  of $G_t$, which coincides with the number of legitimate nodes of the network.

\item $r$ is a large random number.

\item $N_{G_{t}}(v_i)$ denotes the neighbours of node $v_i$ in the graph $G_t$.

\item ${\Pi}(V_t)$ represents a random permutation over the vertex set $V_t$

\item ${\Pi_i}(V_t)$ denotes a random permutation over $V_t$ chosen by $v_i$.

\item ${\Pi}(G_t)$ denotes the graph isomorphic to $G_t$ corresponding to the permutation ${\Pi}(V_t)$.

\item $c \in_r C$ indicates that an element $c$ is chosen at random with uniform distribution from a set $C$.

\item $HC_t$ designates the Hamiltonian cycle used at stage $t$.

\item ${\Pi}(HC_t)$ represents the Hamiltonian cycle $HC_t$ in the graph ${\Pi}(G_t)$.

\item $N_{HC_t}(v_i)$  denotes the neighbours of node $v_i$ in the Hamiltonian cycle $HC_t$.

\item $S$ and $A$ stand for the supplicant and the authenticator, respectively, both during an insertion phase and during the
execution of a ZKP-based access control.

\item $S \rightleftharpoons A$ symbolizes when node $S$ contacts $A$.

\item $A \leftrightarrow S: data $  means that $ A $ and $S$ agree on $data$

\item $A \stackrel{s}{\rightarrow} S: information $  means that $ A $ sends $information$ to $S$ through a secure channel.

\item $A \stackrel{o}{\rightarrow} S: information $  means that $ A $ sends $information$ to $S$ through an open channel.

\item $A \stackrel{b}{\rightarrow} network: information $ represents when $ A $ broadcasts $information$ to all on-line legitimate nodes of the network.

\item $A \stackrel{b}{\leftrightarrow} network: information $ represents a two-step procedure where $ A $
broadcasts $information$ to all on-line legitimate nodes of the network, and receives their answers.

\item $h$ stands for a public hash function.

\item $T$ denotes the threshold length of the off-line period for legitimate nodes.

\end{itemize}

\section{SLCM Scheme}
\label{scheme}

In this section specific details about network initialization, node insertion, access control, and proofs of life and node deletion are given.

\subsection{Initialization}

The set of vertices of the graph corresponds exactly to the set of nodes of the real network during the whole life-cycle of the network. Consequently, the initialization process starts from a set $V_0$ of $n$ vertices corresponding to the nodes of the initial network. Besides, each vertex sub-index may be used as ID (IDentification) for the corresponding node. The first step of the initialization consists of generating jointly and secretly a random permutation $\Pi$  of such a set. The algorithm for generating the cycle $HC_0$ involves three basic steps. First, each node is assigned a different number $v_i$ $\in$ $[1, n]$ according to its IP, then it generates a random permutation  $\Pi_i(V_t)$ and shares it with the other initial nodes through a secure Bluetooth connection. Finally, every node computes the product of all permutation matrices in order to get  $\Pi(V_t)$.  Once this is completed, each legitimate node should know a Hamiltonian cycle $HC_0$ corresponding exactly to such a permutation. The partial graph formed by the edges corresponding to such a Hamiltonian cycle $HC_0$, is completed by adding $n$ groups of $2m/n$ edges, producing the initial edge set $E_0$. Each one of these $n$ groups of edges must have at end-vertex $v_i$, $i=1, 2, ..., n$, and be randomly generated by the node $v^{i}$. The cardinality $2m/n$ of those edge groups must be large enough so that the cardinality of the resulting edge set $|E_0|= m$ guarantees the difficulty of the HCP in $G_0$.

\medskip
%\%%\hline
\smallskip
\textbf{Initialization Algorithm}
%\%%\hline
\medskip

Input: $V_{0}$, with $\left|V_{0}\right| = n$
\begin{enumerate}

      \item    The  $ n $  nodes of the network generate jointly, secretly and randomly the cycle $HC_{0}=\Pi\left(V_{0}\right)$.
      
      \item Each node $ v_i \in V_{0}$ builds the set $N_{G_{0}}(i) =\left\{   \{v_j \in_r V_{0}\} \cup N_{HC_0}(i)\right\}$
      with $\left|N_{G_{0}}(i)\right| = \frac{2m}{n}$.
      
      \item Each node broadcasts $ v_i \stackrel{b}{\rightarrow} network: N_{G_{0}}(i) $
      
      \item  Each node merges $ E_{0}= \bigcup _{i=1,2,...,n} \left\{ (v_i, v_j) : v_j \in N_{G_{0}}(i) \right \}$
      
\end{enumerate}

 Output: $G_{0} = (V_{0}, E_{0})$, with $\left|E_{0}\right| = m $

\smallskip
%\%%\hline
\medskip

\subsection{Insertion}

The insertion phase described in this section works under the assumption of having mutual trust and a secure Bluetooth connection among the authenticator legitimate node $A$ and the supplicant new node $S$. The first step that node $A$ should do is to assign to the new node $S$ the lowest vertex number $v_i$ not assigned to any node in the vertex set $V_t$. This means either using a number previously used by some deleted node or a new number $v_{n+1}$. Afterwards, A should broadcast such an assignment to all on-line legitimate nodes of the network in order to prevent another simultaneous insertion with the same number, and receive their answer. If $A$ receives less than $n/2$ answers, it stops the insertion procedure because the number of nodes aware of the insertion is not large enough. Otherwise, $A$ chooses the corresponding upgrade of the secret Hamiltonian cycle $HC_t$ by selecting at random two neighbor vertices $v_j$ and $v_k$ in order to insert the new node $v_i$ between them, chooses at random a set of $2m/n - 2$ nodes in $V_t$ such that none of them are neighbors in $HC_t$, and broadcasts the set of neighbors $N_{Gt+1}(v_i)$ of $S$ in the new graph $G_{t+1}$ to all on-line legitimate nodes of the network.

\medskip
%\%%\hline
\smallskip
\textbf{Insertion Algorithm}
%\%%\hline
\medskip

Input: At stage $t$ a supplicant node $S$ wants to become a member of the network.
\begin{enumerate}
    \item $S\rightleftharpoons A$ and node $S$ convinces node $A$ to accept its entrance to the network.
    \item $A$ assigns to $S$ the vertex number $v_i$ such that $i = min \{l: v_l \not \in V_t \}$
    \item $A$ broadcasts $ A \stackrel{b}{\leftrightarrow} network: v_i$
    \item \begin {itemize}
    \item If $A$ receives less than $n/2$ answers, she stops the insertion procedure.
    \item Otherwise:
    
     \begin{enumerate}
     \item $A$ chooses at random $\{v_j\in _r V_t,v_k \in _r N_{CH_t}(v_j)\}$
     \item $A$ chooses at random  $N_{G_{t+1}}(v_i)= \{v_j, v_k \} \cup \{ w_1,w_2,..., w_{\frac{2m}{n} -2}  \in _r V_t$ such that
    $\forall w_{l_1},w_{l_2}:    w_{l_1} \not \in N_{CH_t}(w_{l_2})\}$
    \item $A$ broadcasts $ A \stackrel{b}{\rightarrow} network: N_{G_{t+1}}(v_i)$
    \item Each on-line node computes $V_{t+1}=V_t \cup \{v_i\}$, $E_{t+1}=E_t \cup N_{G_{t+1}}(v_i)$ and
     $HC_{t+1}=\{HC_t \setminus (v_j, v_k)\} \cup \{(v_j,v_i) \cup (v_i, v_k)\}$
    \item   $A$ sends openly $A \stackrel{o}{\rightarrow} v_i: G_{t+1}$
    \item  $A$ sends securely $A \stackrel{s}{\rightarrow} v_i: HC_{t+1}$
\end{enumerate}
\end {itemize}
\end{enumerate}

Output: The supplicant node $S$ is a legitimate member of the network.
\smallskip
%\%%\hline
\medskip

\subsection{Access Control}

If a legitimate node $S$ that has been off-line from stage $t$ wants to connect on-line to the network at stage $r$, it first contacts a legitimate on-line member $A$. Afterwards, $A$ should check whether the off-line period of $S$ is not greater than $T$. In this case, $S$ has to be authenticated by $A$ through a ZKP of its knowledge of the secret solution $HC_t$ on the graph $G_t$. The parameter setting of $T$ can be based on the mean time that legitimate nodes of the network have been off-line previously. This value must be regularly updated after each successful access control through the addition of the updated mean and standard deviation plus a positive value epsilon. The initialization of $T$ is done to a value large enough.

\medskip
%\%%\hline
\smallskip
\textbf{Access-Control Algorithm}
%\%%\hline
\medskip
Input: At stage $r$ a supplicant node $S$ that has been off-line from stage $t$ wants to connect on-line to the network.
\begin{itemize}
    \item $S\rightleftharpoons A$
    \item $S$ sends openly $S \stackrel{o}{\rightarrow} A: G_t$
    \item $A$ checks whether       $t \leq r-T$
      \begin {itemize}
      \item if $t \leq r-T$ then $S$ is not authenticated
      \item otherwise:
      \begin{itemize}
        \item $A$ and $S$ agree  $A \leftrightarrow S: l $
        \item $\forall j \in \left\{1,2,\ldots,l\right\}$
        \begin{enumerate}
             \item  $S$  chooses ${\Pi}_j(V_t)$ and builds  ${\Pi}_j(G_t)$ and ${\Pi}_j(HC_t)$, isomorphic graph
             to $G_t$ and correspondent Hamiltonian cycle, respectively
             \item $S$ generates two large random numbers $r_1$ and $r_2$
             \item  $S$ sends openly $S \stackrel{o}{\rightarrow} A: \{ h({\Pi}_j(G_t) || r_1), h({\Pi}_j(HC_t)|| r_2) \}$
             \item $A$ chooses  the challenge $b_j \in_r \left\{0,1\right \} $
             \item $A$ sends openly the challenge  $A \stackrel{o}{\rightarrow} S: b_j$
             \begin{enumerate}                    
                    \item If $b_j=0$ then $S$ sends openly $S \stackrel{o}{\rightarrow} A: \{\Pi_j(G_t),r_1\}$
                    \item If $b_j=1$ then $S$ sends openly $S \stackrel{o}{\rightarrow} A: \{ {\Pi}_j(G_t), {\Pi}_j(HC_t),r_2 \} $
             \end{enumerate}
             \item $A$ verifies
             \begin{enumerate}
                \item that the hash function $h$ on the result of ${\Pi}_j$ on $V_t$ concatenated with $r_1$ produces the value received in step 3, 
                if $b_j=1$
                \item that the hash function $h$ on ${\Pi}_j(HC_t) || r_2$ produces the value received in step 3, and that ${\Pi}_j(HC_t)$ is a valid Hamiltonian cycle in ${\Pi}_j(G_t)$, if $b_j=0$
             \end{enumerate}  
        \end{enumerate}
        \item if $\exists  j \in \left\{1,2,\ldots,l\right\}$ such that the verification is negative, then $S$ is isolated.
        \item otherwise $A$ sends  securely $A \stackrel{s}{\rightarrow} S:$ the necessary information to have full access to protected resources of the network.
    \end{itemize}
    \end{itemize}
\end{itemize}
Output: Node $S$ is connected on-line to the network.
\smallskip
\%%%\hline
\medskip

In the second step of the algorithm, a single commitment scheme based on a cryptographic hash function is used, so that after a random selection of the committed isomorphism, a hash of it and of the isomorphic $HC$ is sent. To open the commitment, $S$ reveals one of those pieces of information thus letting to recalculate the hash and to compare the result with the received hash value.

\subsection{Proofs of Life}

Every on-line legitimate node has to confirm its presence in an active way every certain interval of time of length $T$ through the broadcast of a proof of life. During such a broadcast every node adds its own proof of life to the broadcast so that when the broadcast reaches the last node, a broadcast back starts and when the starting node receives the proofs of life of all on-line legitimate nodes, it rebroadcasts them. Since several nodes might try to broadcast their proofs of life at the same time, in order to reduce such concurrent broadcast, a random timer can be introduced so that each node defers a random time before it sends its proof of life. If it hears another proof of life during this random time, it then gives up its broadcast.

\medskip
\%%\hline
\smallskip
\textbf{Proof-of-Life Algorithm}
\%%\hline
\medskip
Input: At stage $t$ node $A$ is an on-line legitimate node of the network.
\begin{itemize}
    \item $A$ initializes its $clock=0$ just after its last proof of life
    \item if $clock >T$ then
    \begin{enumerate}
\item $A$ broadcasts $ A \stackrel{b}{\leftrightarrow} network: A's$ $ proof$ $of$ $life$
     \item
     \begin {itemize}
     \item If $A$ receives less than $n/2$ proofs of life as answers to its broadcast, it stops its proof of life and puts back its clock.
      \item Otherwise: $A$ broadcasts $ A \stackrel{b}{\rightarrow} network: Received$ $
proofs$ $of$ $life$
\end {itemize}
    \end{enumerate}
    \end{itemize}

Output: At stage $t+1$ node $A$ continues being an on-line legitimate node of the network of the network.

\smallskip
\%%\hline
\medskip

\subsection{Node Deletion}

Each node that has not proven its life is deleted from the network, and the corresponding vertex is deleted from the graph and from the Hamiltonian cycle. This way to proceed guarantees a limited growth of the graph that is used in authentication, and at the same time, allows that always legitimate nodes of the network correspond exactly to vertices in that graph.

\medskip
\%%\hline
\smallskip
\textbf{Deletion Algorithm}
\%%\hline
\medskip
Input: At stage $t$ a node $v_i$ is an off-line legitimate node of the network.
\begin{itemize}
    \item $A$ initializes her $clock=0$
    \item if $clock >T$ then
    \begin{enumerate}
    \item $\forall v_i \in V_t$: $A$ checks $v_i$'s proof of life in $A$'s FIFO queue
    \item $A$ updates $V_{t+1}=V_t \setminus \{v_i\in V_t$ with no proof $\}$
    \item $A$ updates $E_{t+1}=E_t \setminus \{ (v_i, v_j): v_i \in V_t$
    with no proof, $v_j \in N_{G_t(v_i)} \} \cup \{(v_j, v_k): v_j, v_k \in N_{HC_t(v_i)} \}$
        \item $A$ updates $HC_{t+1}=HC_t \setminus \{(v_j, v_i), (v_i, v_k)\} \cup (v_j,v_k): v_i \in V_t$
        with no proof, $v_j, v_k \in N_{HC_t(v_i)}$
    \end{enumerate}
    \item   If $A$ was the starter of the broadcast used for the $v_i$'s deletion, $A$ adds this information to the second step of the
    proof-of-life broadcast:  $A \stackrel{b}{\rightarrow} network: $ $v_i$ is deleted.
\end{itemize}

Output: At stage $t+1$ the node $v_i$ has been deleted both from the network and from the graph.
\smallskip
\%%\hline
\medskip

\section{Security Analysis}
\label{analysis}

In the above sections, several secure algorithms have been presented so that there is no piece of information revealed by any of them that interferes with the security of the others. Thus, the resulting composite protocol is secure. In this section we discuss this issue. For the initialization of the network, there must be a minimum number of nodes to ensure the reliability of the key. Furthermore, these nodes must be legitimate and not be compromised. After initialization, the network will remain working as long as it does not fall below the threshold where the key is no longer safe.

This proposal assumes the ideal environment where all legitimate nodes are honest and where no adversary may compromise a legitimate node of the network in order to read its secret stored information. Such assumption is well suited as a basic model in order to decide under which circumstances the designed authentication scheme is applicable to MANETs. For instance, a possible adaptation of the proposal in order to avoid that hypothesis could be the consideration of a threshold scheme for every step of the scheme, so that every proof of life, insertion, access control or deletion should be done by a group or all nodes each time instead of only one node. In this way, a single dishonest node would not affect the correct operation of the network.

Another requirement of the scheme is the necessary establishment of a secure channel for both the initialization and the insertion procedures where trust between pairs of nodes is assumed. However, that aspect may be easily fulfilled thanks to the fact that most wireless devices can communicate with each other via Bluetooth wireless technology, which however is not valid for general communications because of the short distance it requires.

With respect to possible attacks, due to the lack of a centralized structure, it is natural that possible DoS attacks have the chat application as their main objective. In order to protect the scheme against this threat it must be assured that chat messages, although are publicly readable, may be only sent by legitimate on-line members of the network. Another important aspect related to the use of the chat application is the necessary synchronization of the on-line nodes, so a common network clock is necessary. This requirement has been implemented during simulations through the chat application thanks to the broadcast GRI. The clock can be synchronized with the periodic Proofs of Life. This method is not 100\% accurate, but it has acceptable margins of error.

MANETs are in general vulnerable to different threats such as spoofing and the man-in-the-middle attack. Such attacks are difficult to prevent in environments where membership and network structure are dynamic, and the presence of central directories cannot be assumed. However, our proposal is resistant to spoofing attacks because access control is proved through a ZKP that makes useless the reading of any information published through the chat application or sent openly during an access control. On the other hand, the goal of the man-in-the-middle attack is either to change a sent message or to gain some useful information by one of the intermediate nodes. Again the use of ZKPs in our protocol implies that reading any transferred information does not reveal any useful information about the secret, so changing the message is not possible since only legitimate nodes whose access has been allowed can use the chat application.

Another active attack that might be especially dangerous in MANETs is the so-called Sybil attack. It happens when a node tries to get and use multiple identities. The most extreme case of this type of attacks is the establishment of a false centralized authority who states the identities of legitimate members. However, this specific attack is not possible against our scheme due to its distributed nature. In our scheme, the responsibility of controlling general Sybil attacks will be shared among all the on-line nodes. If an authenticator node detects that a begging node is trying to get access to the network by using an ID that is already being used on-line, such access control must be denied and the corresponding node must be isolated. The same happens when any on-line node detects that an authenticator node is trying to insert a new node to the network with a new ID, and such a node has already assigned a vertex ID. Again, such insertion must be denied and the corresponding supplicant node must be isolated. Anyway, if a Sybil attacker enters the network, any of its neighbours will detect it as soon as it sends proofs of life for different vertex IDs.

\section{Performance Evaluation}
\label{evaluation}

We now analyze the efficiency of the proposal both from the energy consumption and from the computational complexity points of view. We consider the energy consumption, which is the result of transmissions of data and processor activities due to authentication tasks. In the proposal there are two phases when computational overhead is more significant: the ZKP-based access control and the periodic checking of stored FIFO queue. A reduction on the number of rounds of ZKP has a direct effect on the total exchanged messages size in insertions, but a trade-off should be maintained between protocols robustness and performance. Indeed, regarding total data transmission over wireless links, the ZKPs take less than 10\% in a usual situation according to the data we have estimated and obtained in simulations.

The periodic proofs of life accounts for approximately 90\% of the total exchanged message size in many cases. However, we have found that these compulsory proofs of life imply an incentive technique for stimulating cooperation in authentication tasks. This is due to the fact that nodes that are broadcasters of deletions or authenticators in insertions or access controls are exempted from their obligation to broadcast their proofs of life. 

In order to reduce data communication cost of the protocol, an increase on the threshold period $T$ might be an option, but again an acceptable balance should be kept. According to our experiments, $T$ should depend directly on the mean time that nodes are off-line and on the number of legitimate and/or on-line nodes in order to prevent a possible bandwidth overhead of large networks. 

The number of packets generated in the network grows linearly with the number of authenticated nodes on the network. In addition, communications are initiated periodically so although the total number of packets in the network grows, the number of packets in a network area remains nearly constant and this number is only affected if the density in that area is increased.

The energy that a node needs is not affected by the growth of the network, but it is affected by its density. However, the size of storage that a node needs increases as the network grows. This aspect together with the routing problem are the main reasons by which it is necessary an upper limit on the size of the MANET.

For the performance analysis of the proposal we used the Network Simulator NS-2 with DSR routing protocol. We created several Tcl based NS-2 scripts in order to produce various output trace files that have been used both to do data processing and to visualize the simulation. Within our simulation we used the visualization tool of Network Animator NAM and the NS-2 trace files analyzer of Tracegraph. For the simulation of mobility we used the setdest program in order to generate movement pattern files based on the random waypoint algorithm.

\subsection{Hamiltonian Cycle}

\begin{figure}[htb]
  \centering
     \includegraphics[scale=0.15]{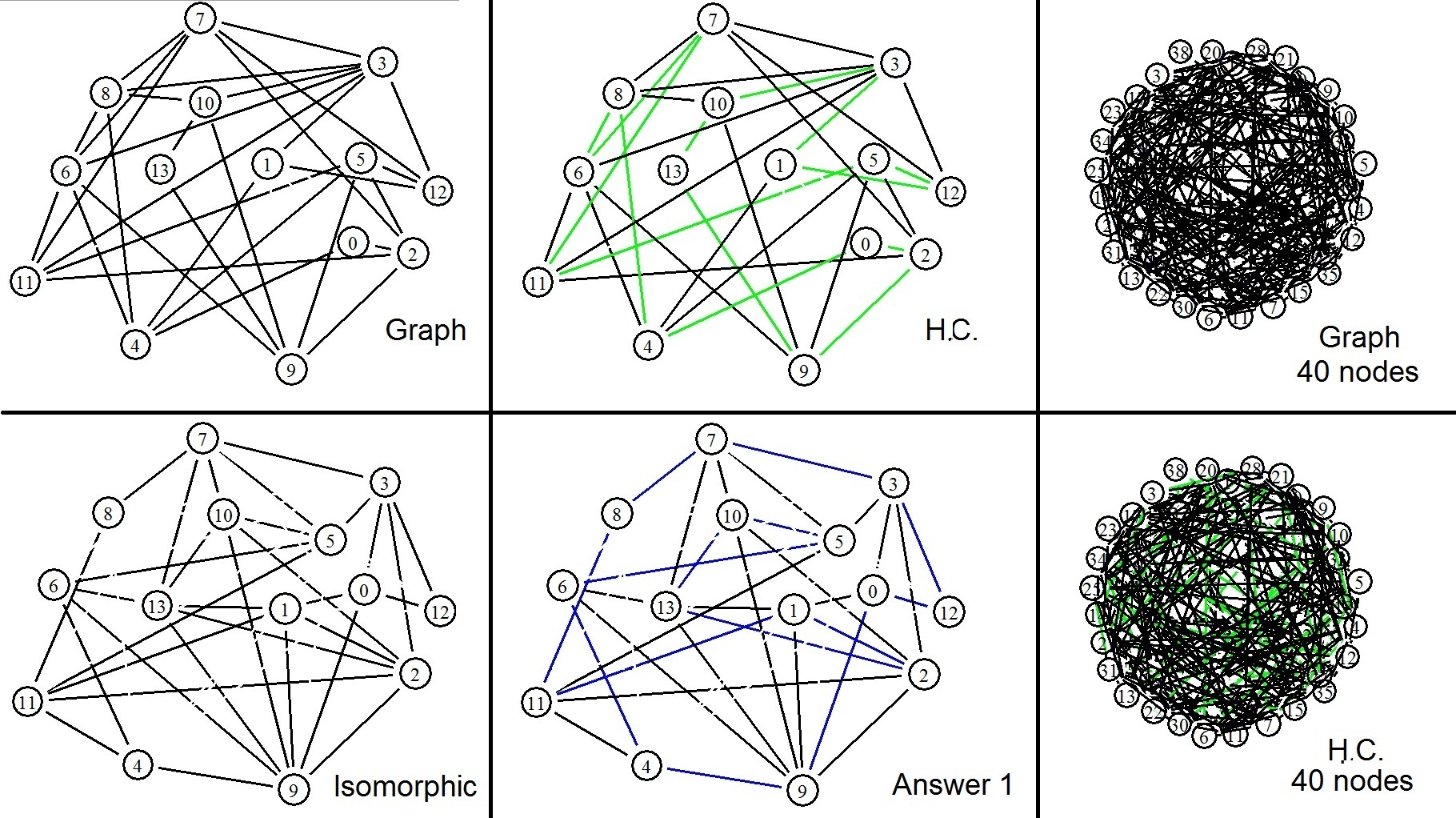}
  \caption{Example of HC-based ZKP}
  \label{fig:Graph}
\end{figure}

Simple examples of a simulation using a few nodes consisting of scenario files that describe the movement patterns of the nodes and communication files that describe the traffic in the network were used to produce trace files that were analyzed to measure various parameters.
The trace files were used to visualize the simulation using NAM, while the measurement values are used as data for plots with Tracegraph. An example of the final graph and Hamiltonian cycle associated to the example network is shown in Figure \ref{fig:Graph} where green is used to indicate the Hamiltonian cycle, blue is used for the inserted nodes and red is used for the edges deleted from the Hamiltonian cycle when inserting new nodes. 

As aforementioned, the HCP is NP-complete. Indeed, searching an HC by backtracking is computationally intractable and the only practical approach is through heuristic algorithms, and even most heuristic algorithms are useless for different types of random graphs with more than 200 nodes \cite{Csehi, Vandergriend}. However, since the protocol does not require solving the problem but constructing a graph from a chosen solution, the difficulty of the HCP is not a disadvantage against the efficiency of the scheme but an advantage in favor of the security of the scheme.

\subsection{Network Topology}

We conducted different simulations to see the effects of different metrics by varying network density and topology. We were changing the number of nodes from 10 to 100, the area from 400 to 1000 $m^2$, and the period of simulation from 60 to 200 seconds. We also used the probabilities of insertions and deletions in each second from 5\% to 25\%, in order to modify the mobility rate and antenna range of nodes from 2 to 15 m/s and 100 to 250 meters respectively. This range also defines different frequencies of accesses to the network.
For the simulation we distinguished different states where nodes can be in the network, depending on different factors:

\begin{itemize}
\item	\textsl{On and Authenticated}: Nodes having no label at the beginning of the simulation or nodes that are labeled \textsl{On} who have gone from \textsl{Off} to On are in this state.

\item	\textsl{Non-Legitimate}: There are nodes that are Off and do not belong to the network. These nodes are candidates to enter the network when they turn on. 

\item	\textsl{On to Authenticate}: When a node is On and asks another node to be authenticated, the node is turned on but still does not belong to the network. Nodes that are turned on but are not authenticated on the network appear to be Off to network effects.

\item	\textsl{Off}: This stage corresponds to the nodes that belong to the network but are off-line. These nodes either can go on and become part of the network after previously demonstration that they know the secret of the network or can be turned off until their period of life ends, in which case the node is removed from the network.

\item	\textsl{Re-Inserted}: When a node in the network is turned off, stays off for less than $T$, and then turns on and shows its knowledge of the network secret to a node that has responded to the last proof of life, it is reactivated.

\item	\textsl{Deleted}: When the node is off-line for a too long time, it goes to this state where it is removed both from the HC and the graph.

\item	\textsl{Out of Service}: A node that is legitimate and on-line but does not respond to a proof of life started by another node because it is unreachable, would have to show that it belongs to the network when it finds another node on the network.

\item	\textsl{Added}: A non-legitimate node that receives enough network information from some legitimate node after an insertion procedure changes its state to \textsl{Added}.

\end{itemize}

To study the performance of the proposed scheme, simulations were performed by using the same density with different numbers of nodes and running time enough to study the effects of the proofs of life by varying only the number of nodes.

From these tests we collected data on the number of connections and the number of generated, forwarded, and lost packets, which are shown in Figure \ref{fig:Generated}

\subsection{Experimental Results}

\begin{figure}[htb]
  \centering
     \includegraphics[scale=0.45]{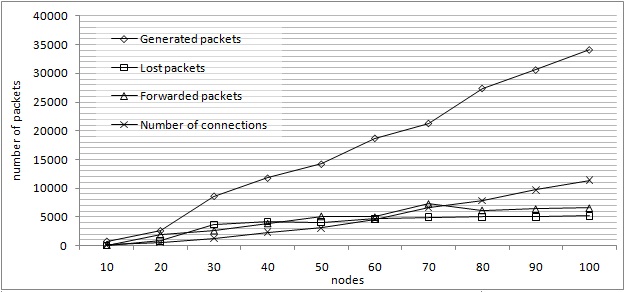}
  \caption{Generated packets}
  \label{fig:Generated}
\end{figure}

This section analyzes different aspects of experimental results, which show the quality and security of the proposed scheme, considering in particular, the relationship with the number of nodes. Figure \ref{fig:Generated} shows that according to simulations of the proposal both the number of connections and the number of generated packets increase linearly with the number of nodes. This happens when the density of the network is maintained by increasing the number of nodes. The picture also shows that the number of forwarded or lost packets also increases with the number of nodes, but in a more contained way than in the case of generated packets. This happens when  the number of nodes and thereby their connections increase, but also the size of the plane increases to maintain a constant density so that the interference between nodes does not vary. Thus, the obtained results regarding lost packets can be considered positive.

\begin{figure}[htb]
  \centering
     \includegraphics[scale=0.55]{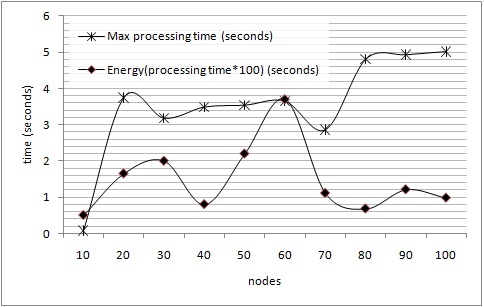}
  \caption{Processing time}
  \label{fig:ProceTime}
\end{figure}

Figure \ref{fig:ProceTime} reflects the average energy and the maximum power consumed by each node. These parameters are calculated from the processing time of packets of each node. This chart allows us to see that the maximum processing time increases with the number of nodes, although there are some exceptions. This is because with a higher density of nodes that initiate the proof of life, more computational work exists in the network. The picture shows that the average processing time is quite low and does not follow a pattern that can be used to relate it to the number of nodes. Anyway, we could conclude that on average the energy consumed by nodes does not increase too much when the network grows.

\begin{figure}[htb]
  \centering
     \includegraphics[scale=0.55]{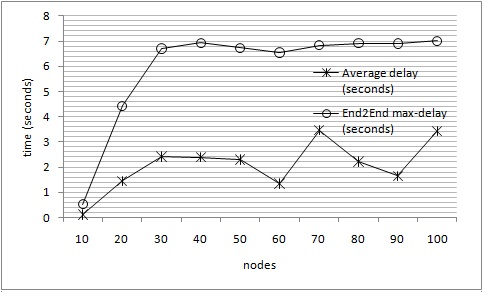}
  \caption{Delay between nodes}
  \label{fig:retardo}
\end{figure}

Figure \ref{fig:retardo} shows both the delay signal between nodes and the biggest delay that occurred in the simulation for different numbers of nodes. In both cases we see a large growth with 30 nodes and then a slight increase. The maximum delay that occurs after the 30 nodes is almost constant in 7 seconds, whereas the average time delay increases to 30 nodes and then fluctuates. These results show a good behavior of the proposal regarding delay of messages produced by the network growth.

\begin{figure}[htb]
  \centering
     \includegraphics[scale=0.50]{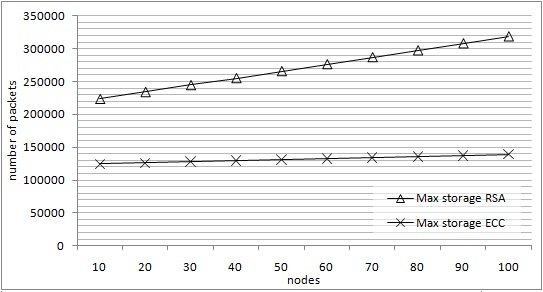}
  \caption{Maximun capacity storage}
  \label{fig:almacenamiento}
\end{figure}

Figure \ref{fig:almacenamiento} shows the need for maximum storage required for each node in bits. Indeed, since the proposal does not require almost any storage, the shown growth is because each node can need to store the public keys and other data of the remaining nodes in the network. Also each node could store a number of certificates signed by and for the other nodes to authenticate them. We compared the need for storage using 1024-bit keys for RSA and 160 bits for Elliptic Curve Cryptography (ECC), which are cryptographically equivalent.

Thus, some conclusions that we can deduce from the simulations are:

\begin{itemize}
\item	The protocol scales to any sort of networks with different levels of topology changes.

\item	Node density is a key factor for the mean time of insertions, but it is not as large as it might be assumed.

\item	$A$ right choice of parameter $T$ should be done according to mean off-line time, number of nodes, bandwidth of wireless connections and computation and storing capacities of nodes.

\item	$A$ positive aspect of the proposal is that the requirements in the device hardware are very low.
\end{itemize}

\section{Conclusion}

A complete self-organizing life cycle management scheme for MANETs called SLCM has been designed and simulated in this work. Unlike conventional networks, MANETs have no infrastructure and therefore their management must be self-organized, which is a major research challenge due to the security needs of communications. This work describes a new fully self-organized authentication scheme that has been specially designed for MANETs, and that supports knowledge-based member authentication in server-less environments. The overall goal of the proposal has been to design a strong authentication scheme able to react and adapt to network topology changes without the necessity of any centralized authority. Its core technique consists of a Zero-Knowledge Proof to avoid the transference of any relevant information. The proposal is balanced because the procedures that legitimate members of the network have to carry out when the network is upgraded (insertion or deletion of nodes) imply identical work for every legitimate node. The development and evaluation of many NS-2 simulations of the proposal is an important part of this work. The obtained results show the scalability and robustness of the proposal.

\section*{Acknowledgement} Research supported by the Ministerio de Ciencia e Innovaci\'on and the European FEDER Fund under Project TIN2008-02236/TSI and FPI scholarship BES-2009-016774, and by the Agencia Canaria de Investigaci\'on, Innovaci\'on y Sociedad de la Informaci\'on under PI2007/005 Project and FPI scholarship BOC Number 60.


\begin{thebibliography}{9}

\bibitem{Aboudagga} Aboudagga~N, Tamer~M, Eltoweissy~M,  DaSilva~L,  Quisquater~J. 2005. \emph{Authentication protocols for ad hoc networks, Taxonomy and research Issues}. 1st ACM international workshop on Quality of service and security in wireless and mobile networks, 300-311.

\bibitem{Azer}	Azer~MA, El-Kassas~SM, El-Soudani~MS. 2007. \emph{Threshold Cryptography and Authentication in Ad Hoc Networks Survey and Challenges}. Second International Conference on Systems and Networks Communications.

\bibitem{Badonnel} Badonnel~R. 2007. \emph{Management of Ad-Hoc Networks and Services}. Thesis, LORIA - INRIA.

\bibitem{Bonnefoi} Bonnefoi~PF, Sauveron~D, Park~JH. 2008. \emph{MANETS: an exclusive choice between use and security?}. Computing and Informatics. Vol. 27, no. 5, 799-821.

\bibitem{CCMQ} Caballero-Gil~P, Caballero-Gil~C, Molina-Gil~J, Quesada-Arencibia~A. 2007. \emph{A Simulation Study of New Security Schemes in Mobile Ad-hoc NETworks}. Lecture Notes in Computer Science. Vol. 4739. 217-224.

\bibitem{Caballero} Caballero-Gil~P, Hern\'andez-Goya~C. 2006. \emph{Zero-Knowledge Hierarchical Authentication in MANETs}. IEICE Transactions on Information and Systems. Vol. E-89-D. 1288-1289.

\bibitem{Chang} Chang~J, Yang~H, Chao~H, Chen~J. 2010. \emph{Multipath design for 6LoWPAN ad hoc on-demand distance vector routing}. International Journal of Information Technology, Communications and Convergence. Vol.1. , No.1, 24-40.

\bibitem{Csehi} Csehi~C, T\'oth~J. 2011. \emph{Search for Hamiltonian Cycles}. The Mathematical Journal, Vol.13. 1-16.

\bibitem{Freudiger}	Freudiger~J, Raya~M, Hubaux~JP. 2009. \emph{Self-organized Anonymous Authentication in Mobile Ad Hoc Networks}. Lecture Notes of the Institute for Computer Sciences, Social Informatics and Telecommunications Engineering. Vol. 19, Part 7, 350-372.

\bibitem{Goldreich}	Goldreich~O, Micali~S, Wigderson~A. 1986. \emph{How to prove all NP-statements in zero-knowledge, and a methodology of cryptographic protocol design}. Crypto '86, Lecture Notes in Computer Science. Vol. 263. Springer-Verlag. 171-185.

\bibitem{Hahm}	Hahm~S, Jung~Y, Yi~S, Song~Y, Chong~I, Lim~K. 2005. \emph{A Self-Organized Architecture in Mobile Ad-Hoc Networks}. ICOIN, Lecture Notes in Computer Science. Vol. 3291. Springer-Verlag. 689-696.

\bibitem{Hubaux}	Hubaux~JP, Butty\'an~L, Capkun~S. 2001. \emph{The quest for security in mobile ad hoc networks}. The ACM International Symposium on Mobile Ad Hoc Networking and Computing. MobiHoc. 146-155.

\bibitem{Imani} Imani~M, Taheri~M, Naderi~M. 2010. \emph{Security enhanced routing protocol for Ad hoc networks}. Journal of Convergence Security enhanced routing protocol for ad hoc networks. Volume 1, Number 1. 43-48.

\bibitem{Kambourakis}	Kambourakis~G, Konstantinou~E, Douma~A, Anagnostopoulos~M, Fotiadis~G. 2010. \emph{Efficient Certification Path Discovery for MANET}. EURASIP Journal on Wireless Communications and Networking. Vol. 2010. 1-16.

\bibitem{Luo} Luo~J, Hubaux~JP, Eugster~PT. 2005.  \emph{DICTATE: DIstributed CerTification Authority with probabilisTic frEshness for Ad Hoc Networks}. IEEE Transactions on Dependable Secure Computing, Vol. 2(4), 311-323, IEEE.

\bibitem{Maki}	Maki~S, Aura~T, Hietalathi~M. 2000. \emph{Robust membership management for ad-hoc groups}. 5th Nordic Workshop on Secure IT Systems NORDSEC.

\bibitem{Prahmkaew} Prahmkaew~S. 2010.  \emph{Performance Evaluation of Convergence Ad Hoc Networks}. Journal of Convergence. Vol.1, No.1, 101-106.

\bibitem{Saxena}	Saxena~N, Tsudik~G, Yi~JH. 2005. \emph{Efficient node admission for short-lived mobile ad hoc networks}. IEEE International Conference on Network Protocols ICNP. 269-278.

\bibitem{Stajano}	Stajano~F, Anderson~R. 1999. \emph{The Resurrecting Duckling: Security Issues for Ad-Hoc Wireless Networks}. 7th International Workshop on Security Protocols. 172-294.

\bibitem{Tseng}	Tseng~YM. 2007. \emph{A secure authenticated group key agreement protocol for resource-limited mobile devices}. The Computer Journal. Vol. 50(1). 41-52.

\bibitem{Vandergriend} Vandergriend~B. 1998. \emph{Finding Hamiltonian Cycles: Algorithms, Graphs and Performance}. Masters Thesis. University of Alberta.

\bibitem{Xie} Xie~B, Kumar~A, Zhao~D, Reddy~R, He~B. 2010. \emph{On secure communication in integrated heterogeneous wireless networks}. International Journal of Information Technology, Communications and Convergence. Vol.1. , No.1, 4-23.

\bibitem{Yang} Yang~Y, Chen~J, Duan~L, Meng~L, Gao~Z, Qiu~X. 2009. \emph{A self-configuration management model for clustering-based MANETs}. Proc. International Conference on Ultra Modern Technology. 1-7.
 
\bibitem{Zhou} Zhou~L, Haas~J. 1999. \emph{Securing Ad Hoc Networks}. IEEE Networks Special Issue on Network Security. Vol 13 Issue:6. 24-30.

\end{thebibliography}
\end{document}